\documentclass[twocolumn,floats,floatfix,superscriptaddress,aps,pra]{revtex4}
\usepackage{amsmath}
\usepackage{graphicx}
\usepackage{multirow}
\usepackage{booktabs}
\usepackage{afterpage}

\usepackage{array}
\newcolumntype{L}[1]{>{\raggedright\let\newline\\\arraybackslash\hspace{0pt}}m{#1}}
\newcolumntype{C}[1]{>{\centering\let\newline\\\arraybackslash\hspace{0pt}}m{#1}}
\newcolumntype{R}[1]{>{\raggedleft\let\newline\\\arraybackslash\hspace{0pt}}m{#1}}

\def\be{ \begin{equation} }
\def\ee{ \end{equation} }

\def\A{A}

\def\to{\rightarrow}

\def\matrix22#1#2#3#4{\left[ \begin{array}{cc} #1 & #2 \\ #3 & #4 \end{array}\right]}

\def\bt{\begin{tabular}}
\def\et{\end{tabular}}

\begin{document}

\author{Boyan T. Torosov}
\affiliation{Institute of Solid State Physics, Bulgarian Academy of Sciences, 72 Tsarigradsko chauss\'{e}e, 1784 Sofia, Bulgaria}
\author{Nikolay V. Vitanov}
\affiliation{Department of Physics, St Kliment Ohridski University of Sofia, 5 James Bourchier Blvd, 1164 Sofia, Bulgaria}

\title{Experimental demonstration of composite pulses on IBM's quantum computer}

\date{\today}

\begin{abstract}

We perform comprehensive experimental tests of various composite pulse sequences using one of open-access IBM's quantum processors, based on superconducting transmon qubits. 
We implement explicit pulse control of the qubit by making use of the opportunity of low-level access to the backend, provided by IBM Quantum.  
We obtain the excitation profiles for a huge variety of broadband, narrowband, and passband composite pulses, producing any pre-chosen target probabilities, ranging from zero to one.
We also test universal composite pulses which compensate errors in any experimental parameter.
In all experiments, we find excellent agreement between theoretical and experimental excitation profiles.
This proves both the composite pulses as a very efficient and flexible quantum control tool and the high quality of the IBM quantum processor.
As an extreme example, we test and observe a pronounced narrowband excitation profile for a composite sequence of as many as 1001 pulses.

\end{abstract}

\maketitle

\section{Introduction}\label{sec-introduction}

Composite pulses (CPs) are a powerful quantum control technique, used in a wide range of different physical systems and control tasks. 
CPs were first developed and used in nuclear magnetic resonance imaging (NMR) \cite{NMR, Levitt1986, Levitt2007} as a convenient tool for robust manipulation of spins by magnetic fields. 
A composite pulse consists of a series of single pulses each having a well-defined phase. 
With a proper selection of the phases of the ingredient pulses, the composite pulse can compensate for fairly large systematic errors in the driving fields (e.g. miscalibrated or shifted intensity or frequency), which usually lead to poor performance of a single pulse from the sequence. 
After their successful application in NMR, composite pulses have been applied in other areas where robust and highly accurate coherent quantum control is needed.
Applications include qubit control in trapped ions \cite{Gulde2003, Schmidt-Kaler2003, Haffner2008, Timoney2008, Monz2009, Shappert2013, Mount2015, Zarantonello2019}, neutral atoms  \cite{Rakreungdet2009,Demeter2016}, quantum dots \cite{Wang2012,Kestner2013,Wang2014,Zhang2017,Hickman2013,Eng2015}, NV centers in diamond \cite{Rong2015}, doped solids \cite{Schraft2013,Genov2017,Bruns2018,Genov2014}, superconducting phase qubits \cite{SteffenMartinisChuang},
 optical clocks \cite{Zanon-Willette2018}, atom optics \cite{Butts2013,Dunning2014,Berg2015}, magnetometry \cite{Aiello2013}, optomechanics \cite{Ventura2019}, etc.

%
CPs offer the unique capability to shape the excitation profile in essentially any desired manner, and hence remove the control limitations of a single resonant pulse.
For example, the most ubiquitous CPs are the broadband composite $\pi$ pulse, which produce unit transition probability not only for a pulse area $\A=\pi$ and exact resonance, as a single resonant $\pi$ pulse, but also in some broad ranges around these values \cite{Levitt1979, Freeman1980, Levitt1982, Levitt1983, Levitt1986, Wimperis1990, Wimperis1991, Wimperis1994, Levitt2007, Torosov2011PRA,Torosov2011PRL, Schraft2013,Genov2014}.
In this manner, a sequence of low-quality pulses can produce the performance of a perfect $\pi$ pulse, i.e. a perfect X gate.
At the other extreme, narrowband (NB) composite pulses \cite{Tycko1984,Tycko1985, Shaka1984, Wimperis1990,Wimperis1994, Torosov2011PRA, Vitanov2011,Ivanov2011,Merrill2014, Torosov2020} can dramatically squeeze the excitation profile around the value $\pi$ of the pulse area (or any other chosen value): they produce significant excitation only for a narrow range of pulse areas and suppress excitation outside this range via destructive interference. 
These CPs have interesting applications to sensing, metrology, spatial localization and elimination of cross talk between neighboring qubits.
A very interesting type of CPs are the passband (PB) pulses, which combine the advantages of BB and NB pulses: highly accurate excitation inside a certain parameter range and negligibly small excitation outside it \cite{Cho1986,Cho1987,Wimperis1989, Wimperis1994, Ivanov2011, Kyoseva2013}. 
They can be seen as NB pulses with a BB center of the excitation profile and have similar applications as NB pulses with the added benefit of some robustness around the center.
For instance, they can be used to both reduce the cross talk and simultaneously mitigate the pointing instability of laser driven ions or atoms. 
These three types of CPs have been extended to half-$\pi$ and generally $\theta$ pulses \cite{Levitt1986, Wimperis1994, Levitt2007,Torosov2019variable,Torosov2020}, which produce excitation profiles ``locked'' at any pre-selected value of the transition probability. Half-$\pi$ pulses, in particular, can be used for robust high-fidelity Hadamard gates.
Moreover, a number of CPs have been designed, which compensate variations in the detuning \cite{Levitt1986, Levitt2007}, and simultaneous variations in the pulse area and the detuning \cite{Levitt1986, Levitt2007}.
Recently, universal CPs \cite{Genov2014} have been presented, which compensate deviations in all experimental parameters.

In this work, we put to the test a number of previously derived composite sequences using a publicly available superconducting qubit. 
We use ibmq\_armonq, which is one of the open-access IBM Quantum Canary Processors, and consists of a single transmon qubit. 
We perform experiments to obtain the excitation profiles of BB \cite{Torosov2011PRA}, NB \cite{Vitanov2011}, PB \cite{Kyoseva2013}, and universal CPs \cite{Genov2014}, as well as \emph{theta pulses} \cite{Torosov2019variable,Torosov2020}. 
We selected the ibmq\_armonq because it provides free access to a low-level pulse control of the qubit. This allows for a much deeper study of the accuracy and robustness of the control methods, than the typically used high-level gate-programming layer.
Furthermore, superconducting qubits are leading the race for building large-scale quantum computers and composite pulses might be very useful in boosting qubit fidelity and enhancing resistance to variations in the experimental controls. 

The paper is arranged as follows.
In Sec.~\ref{sec-experiment} we describe the experiment on the IBM qubit.
In Sec. \ref{sec-CP} we present the experimental results for the excitation profiles of the BB, NB, PB and universal CPs derived in \cite{Torosov2011PRA, Vitanov2011, Kyoseva2013, Genov2014}, which produce complete population transfer.
Then, in Sec.~\ref{sec-Theta} we show experimental data on the composite $\theta$ pulses, which produce partial population transfer with the transition probability $\sin^2\theta/2$.
Finally, the conclusions are presented in Sec.~\ref{sec-discussion}.

\section{Description of the experiment}\label{sec-experiment}

The experiments, producing the excitation profiles of the composite sequences in this work, are performed using IBM Quantum Experience \cite{ibm_quantum}. The processor used is ibmq\_armonk v2.4.26, which is one of the IBM Quantum Canary Processors. It consists of a single transmon qubit, controlled by microwave pulses, using the low-level quantum computing module Qiskit Pulse \cite{QiskitPulse}, part of the open-source framework Qiskit \cite{Qiskit}. The parameters of the qubit, calibrated at the time of the experiment, are the following. The qubit has a transition frequency of 4.972 GHz and anharmonicity of -0.34719 GHz. The T1 and T2 coherence times are 203.44 $\mu$s and 301.91 $\mu$s, and the readout error is 3.57\%.

For each performed experiment we apply a sequence of rectangular pulses with the appropriate phase shifts, where each single pulse has a duration of $T=100$ ns. The Rabi frequency, corresponding to a perfect $\pi$ rotation, is therefore $\Omega = \pi/T=2\pi\times 5$ MHz. In our figures we have plotted the transition probability as a function of the deviation $\epsilon$ from this perfect value, $\Omega\to \Omega(1+\epsilon)$, where $\epsilon$ takes values in the range from $-1$ to 1. Every experiment is repeated 1024 times and the results are averaged to obtain a single data point for our plots.

We note here that in the figures, presenting the experimental results, the excitation profiles do not reach the desired target probability and differ from the theoretical predictions with a small amount. This is largely due to the measurement error, which is on the order of 3.5\%.

\section{Composite pulses for complete population transfer}\label{sec-CP}

We first examine three classes of composite sequences, producing complete population transfer over a BB \cite{Torosov2011PRA}, NB \cite{Vitanov2011}, and PB \cite{Kyoseva2013} excitation profiles versus deviations in the Rabi frequency. 
Next, we test the universal CPs \cite{Genov2014}, which provide robust population transfer with respect to all kinds of errors, as long as the evolution is unitary. 
All composite sequences in this section assume an odd total number $N$ of imperfect but identical pulses with a nominal (i.e. in the absence of errors) pulse area of $\pi$.

\subsection{Broadband pulses}\label{sec-BB}

\begin{figure}[tb]
	\includegraphics[width=0.8\columnwidth]{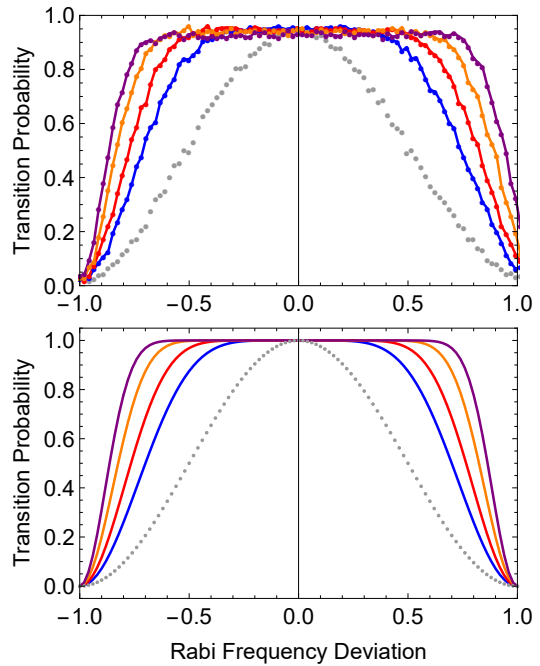}
	\caption{Excitation profiles for the broadband composite pulses, with the phases of Eq.~\eqref{phasesBB}, for $N=3,5,9,15$ (from inside to outside). The gray dotted curve depicts the single-pulse profile as a reference. The upper frame demonstrates the experimental results, while the lower frame shows the theoretical predictions.}
	\label{Fig:BB}
\end{figure}

We start with testing the BB CPs, derived in \cite{Torosov2011PRA}. 
These sequences of nominal $\pi$ pulses with relative phases, used as free parameters, are derived by canceling derivative terms in the propagator expansion vs Rabi frequency error at the zero error point. 
The composite phases $\phi_k$, derived by this approach, read \footnote{This equation is an equivalent, but simpler alternative to the formula (9), derived in \cite{Torosov2011PRA}} 
\be\label{phasesBB}
\phi_k = k(k-1)\pi/N\qquad (k=1,\ldots,N).
\ee
The excitation profiles of these sequences are shown in Fig.~\ref{Fig:BB}, for sequences with $N=3,5,9,15$ pulses. As seen from the figure, the agreement between theoretical predictions and experimental data is excellent.
As the number of pulses in the BB sequence increases, the excitation profiles get broader and broader.

\subsection{Narrowband pulses}\label{sec-NB}

\begin{figure}[tb]
	\includegraphics[width=0.8\columnwidth]{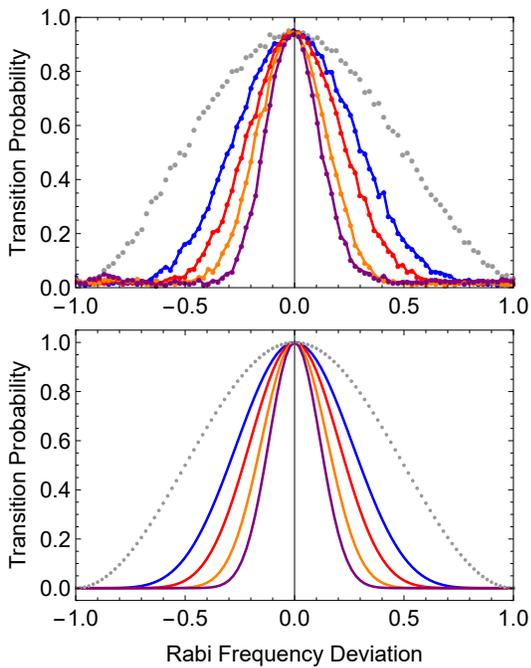}
	\caption{Excitation profiles for the NB composite pulses with the phases of Eq.~\eqref{phasesNB}, for $N=3,5,9,15$ (from outside to inside). The gray dotted curve depicts the single-pulse profile as a reference. The upper frame demonstrates the experimental results, while the lower frame shows the theoretical predictions.}
	\label{Fig:NB}
\end{figure}

Using a similar approach as for the BB pulses, but cancelling the derivative terms in the propagator expansion vs Rabi frequency error at zero Rabi frequency (rather than at Rabi frequency corresponding to a $\pi$ pulse), sequences with NB profiles have been derived in \cite{Vitanov2011}. The phases of these CPs are 
\be\label{phasesNB}
\phi_k = 
\left\{ \begin{array}{c} 
k\pi/N \qquad (k=2,4,\ldots,N-1)   \\ 
-(k-1)\pi/N \qquad (k=1,3,\ldots,N)  
\end{array}\right. .
\ee
The excitation profiles for NB sequences of $N=3,5,9,15$ pulses are illustrated in Fig.~\ref{Fig:NB}. 
As seen from the figures, NB CPs perform as expected, in an excellent agreement with the theory. 
NB pulses can be very useful for spatial localisation and elimination of unwanted cross talk to neighboring qubits.

\begin{figure}[tb]
	\includegraphics[width=0.8\columnwidth]{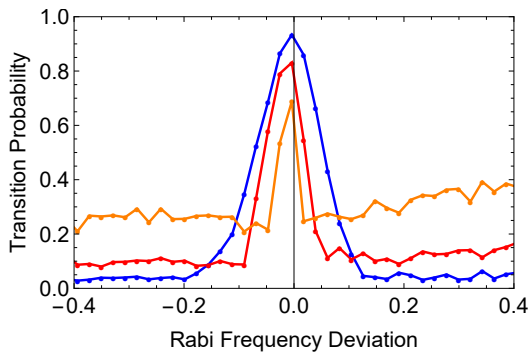}
	\caption{Excitation profiles for the NB composite pulses with the phases of Eq.~\eqref{phasesNB}, for extreme numbers of constituent pulses, $N=75,225,1001$ (from outside to inside).}
	\label{Fig:NBextreme}
\end{figure}

We now push the limits in terms of number of pulses, for which the analytical formula for the phases \eqref{phasesNB} presents an ample opportunity. We measured a few extreme cases with $N=75,225,1001$, which are shown in Fig.~\ref{Fig:NBextreme}. 
To the best of our knowledge, such long CPs have not been used in an experiment hitherto. 
Much to our surprise, the NB sequences perform excellently even for such \textit{huge} numbers of pulses, with clearly squeezing excitation profile.
We note that longer sequences lead to a reduction of the peak transition probability and a raise of the wings, which is an expected behaviour due to the dephasing of the qubit.
Indeed, the 1001-pulse sequence has a total duration of about 100 $\mu$s, which is about a half of the coherence times $T1$ and $T2$.
We also note that the ability of the IBM qubit to maintain such outrageously huge CPs, the NB effect of which crucially depends on quantum coherence, demonstrates the very high quality of this qubit.

\subsection{Passband pulses}\label{sec-PB}

\begin{figure}[tb]
	\includegraphics[width=0.8\columnwidth]{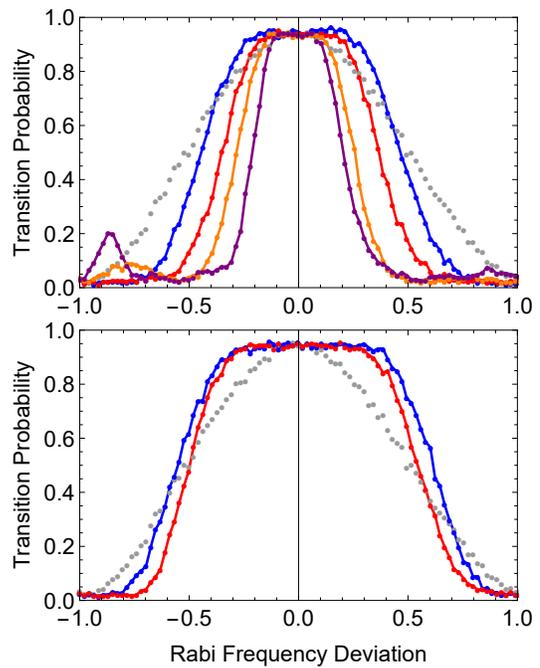}
	\caption{Excitation profiles for the passband composite pulses, derived in \cite{Kyoseva2013}, for the sequences $B_3(N_3)$, $B_3(N_5)$, $B_3(N_9)$, $B_3(N_{15})$ (upper frame, from outside to inside) and $N_3(B_3)$ and $N_5(B_3)$ (bottom frame, from outside to inside). The gray dotted curve depicts the single-pulse profile as a reference.}
	\label{Fig:PB}
\end{figure}

By combining BB and NB pulses one can construct PB composite sequences, as demonstrated in \cite{Kyoseva2013}. 
Two types of such CPs have been derived, named $\mathcal{N}(\mathcal{B})$ and $\mathcal{B}(\mathcal{N})$, meaning nesting a BB pulse into a NB pulse, or a NB pulse into a BB pulse, respectively. 
The first type can be written as
\be
\mathcal{N}(\mathcal{B}) = \{\mathcal{B}_{\nu_1},\mathcal{B}_{\nu_2},\ldots,\mathcal{B}_{\nu_{N_n}}\}
\ee
and corresponds to a NB sequence (with phases $\nu_k$) of $n$ BB sequences $\mathcal{B}$ replacing the nominal $\pi$ pulses in an ordinary NB sequence.
The second type
\be
\mathcal{B}(\mathcal{N}) = \{\mathcal{N}_{\beta_1},\mathcal{N}_{\beta_2}^{-1},\mathcal{N}_{\beta_3},\mathcal{N}_{\beta_4}^{-1},\ldots,\mathcal{B}_{\beta_{N_b}}\}
\ee
stands for a BB sequence (with phases $\beta_k$) of NB sequences $\mathcal{N}$ and their reverse $\mathcal{N}^{-1}$. 

In Fig.~\ref{Fig:PB} we plot the excitation profiles for the $B_3(N_3)$, $B_3(N_5)$, $B_3(N_9)$, $B_3(N_{15})$ sequences (upper frame), as well as the $N_3(B_3)$ and $N_5(B_3)$ sequences (bottom frame). All tested sequences perform as expected and provide a nice PB type of excitation profile.
The measured PB profiles are in excellent agreement with the theoretical ones which are omitted for the sake of brevity.

\begin{figure*}[tbph]
	\includegraphics[width=2\columnwidth]{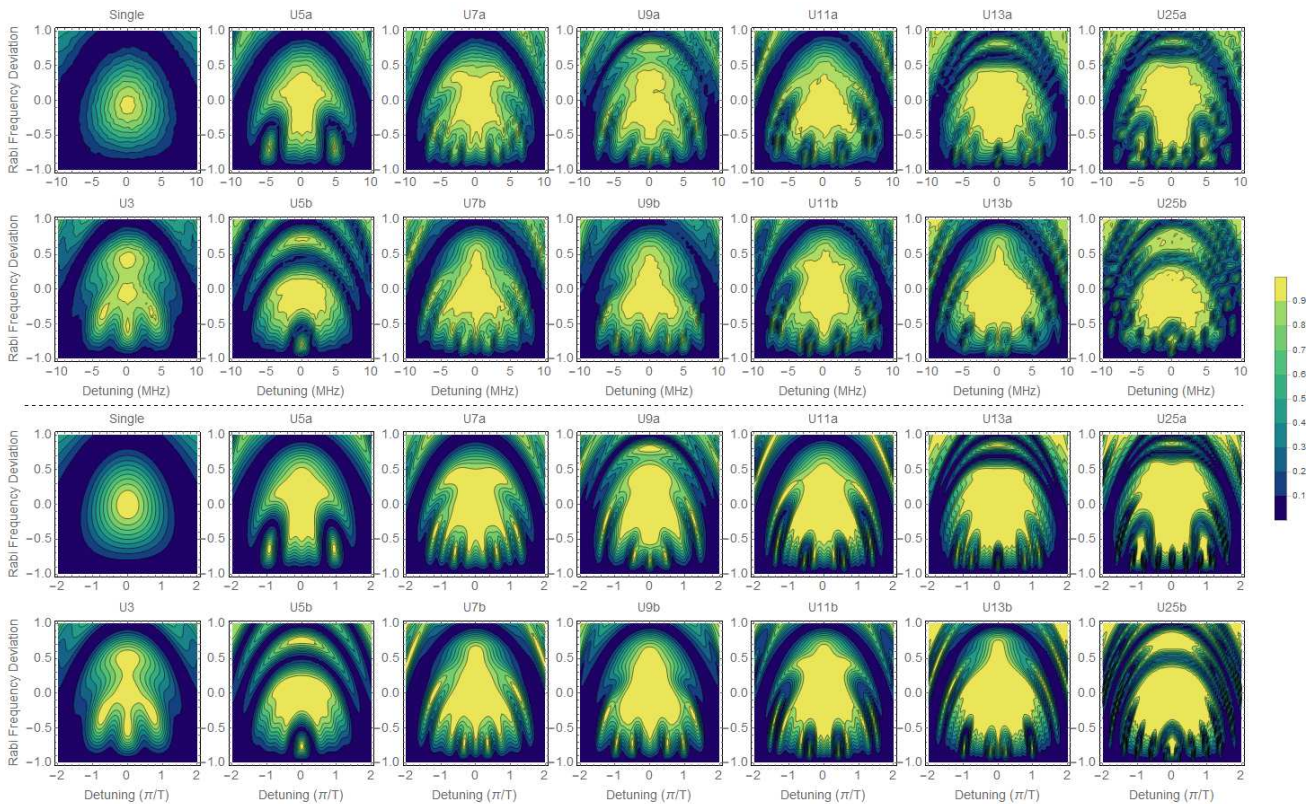}
	\caption{Transition probability as a function of Rabi frequency and detuning for universal composite sequences with phases, given in Table~\ref{table:uni_phases}. The upper part demonstrates the experimental results, while the lower part shows the theoretical predictions.}
	\label{Fig:UniCP}
\end{figure*}

Of particular interest are the $\mathcal{B}(\mathcal{N})$ profiles (upper frame), which are much narrower.
They can be seen as NB profiles (Fig.~\ref{Fig:NB}) with a flatter top.
Hence they can be used to both suppress unwanted cross talk to neighboring qubits and mitigate laser beam pointing instability in laser-controlled ions and atoms.

\subsection{Universal composite pulses}\label{sec-UniCP}

\begin{table}[bt]
\caption{Phases of universal CPs. } 
\begin{tabular}{ll} 
\hline 
Pulse & Phases \\ 
\hline 
 U3 & $(0, 1,0)\pi/2$ \\
U5a & $(0, 5, 2,5,0)\pi/6$ \\
 U5b & $(0, 11, 2,11,0)\pi/6$ \\
U7a & $(0, 11, 10, 17,10,11,0)\pi/12$ \\
 U7b & $(0, 1, 14, 19, 14, 1, 0)\pi/12$ \\
U9a & $(0, 0.366, 0.638, 0.435, 1.697, 0.435, 0.638, 0.366, 0)\pi$ \\
 U9b & $(0, 0.634, 1.362, 0.565, 0.303, 0.565, 1.362, 0.634, 0)\pi$ \\
U11a & $(0, 11, 10, 23, 1, 19, 1, 23, 10, 11, 0)\pi/12$ \\
U11b & $(0, 1, 14, 13, 23, 17, 23, 13, 14, 1, 0)\pi/12$ \\
U13a & $(0, 9, 42, 11, 8, 37, 2, 37, 8, 11, 42, 9, 0)\pi/24$ \\
U13b & $(0, 33, 42, 35, 8, 13, 2,13,8,35,42,33,0)\pi/24$ \\
U25a & $(0, 5, 2, 5, 0, 11, 4, 1, 4, 11, 2, 7, 4,7,2,11,4,1,4,11,$ \\ & $0,5,2,5,0)\pi/6$ \\
U25b & $(0, 11, 2, 11, 0, 5, 4, 7, 4, 5, 2, 1, 4,1,2,5,4,7,4,5,$ \\ & $0,11,2,11,0)\pi/6$ \\ 
\hline 
\end{tabular}
\label{table:uni_phases} 
\end{table}

The universal CPs, derived in \cite{Genov2014}, have the unique ability to compensate all kinds of systematic errors in the experimental parameters, as long as the evolution is governed by a Hermitian Hamiltonian, i.e. is coherent. 
The composite phases of the tested universal sequences are given in Table~\ref{table:uni_phases}. 
The corresponding excitation profiles as a function of the Rabi frequency and the detuning are plotted in Fig.~\ref{Fig:UniCP}.
A remarkable agreement is observed once again between experiment (upper frames) and theory (lower frames), even in the fine details, which we found truly fascinating.
The figure demonstrates the robust excitation, achieved by the universal composite pulses, clearly visible by the broad high-excitation areas, which grow broader as the number of constituent pulses in the sequences increase from 3 to 13. The slight decrease of efficiency for 25-pulse sequences is attributed to decoherence which starts to play a role for such long sequences.

\section{Theta pulses}\label{sec-Theta}

In this section we test composite pulse sequences which generate any pre-selected transition probability with broadband, narrowband, or passband profiles \cite{Torosov2019variable, Torosov2020}. 
For all tested $\theta$ pulses we have chosen the target probabilities to range from $p=0.1$ to $p=0.9$, with a step of $0.1$.

\begin{figure}[t!]
	\includegraphics[width=0.95\columnwidth]{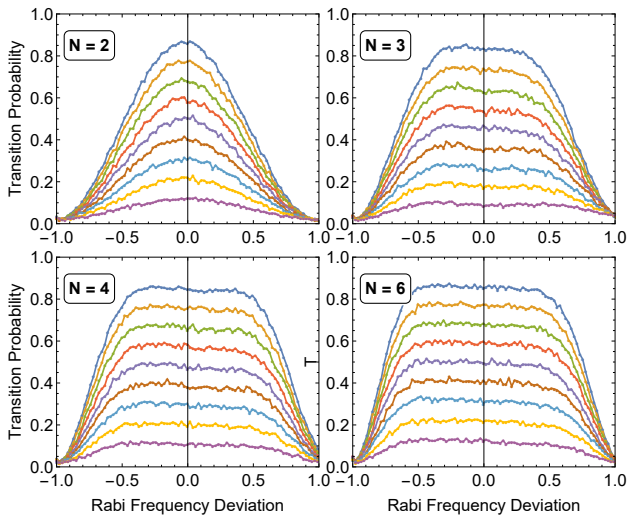}
	\caption{Transition probability induced by composite BB theta pulses with lengths of $N=2,3,4,6$. The composite phases are given in Table~\ref{Table:ThetaBB} and the transition probability is locked at levels $0.1,0.2,\ldots,0.9$.}
	\label{Fig:ThetaBB}
\end{figure}

\begin{figure}[t!]
	\includegraphics[width=0.95\columnwidth]{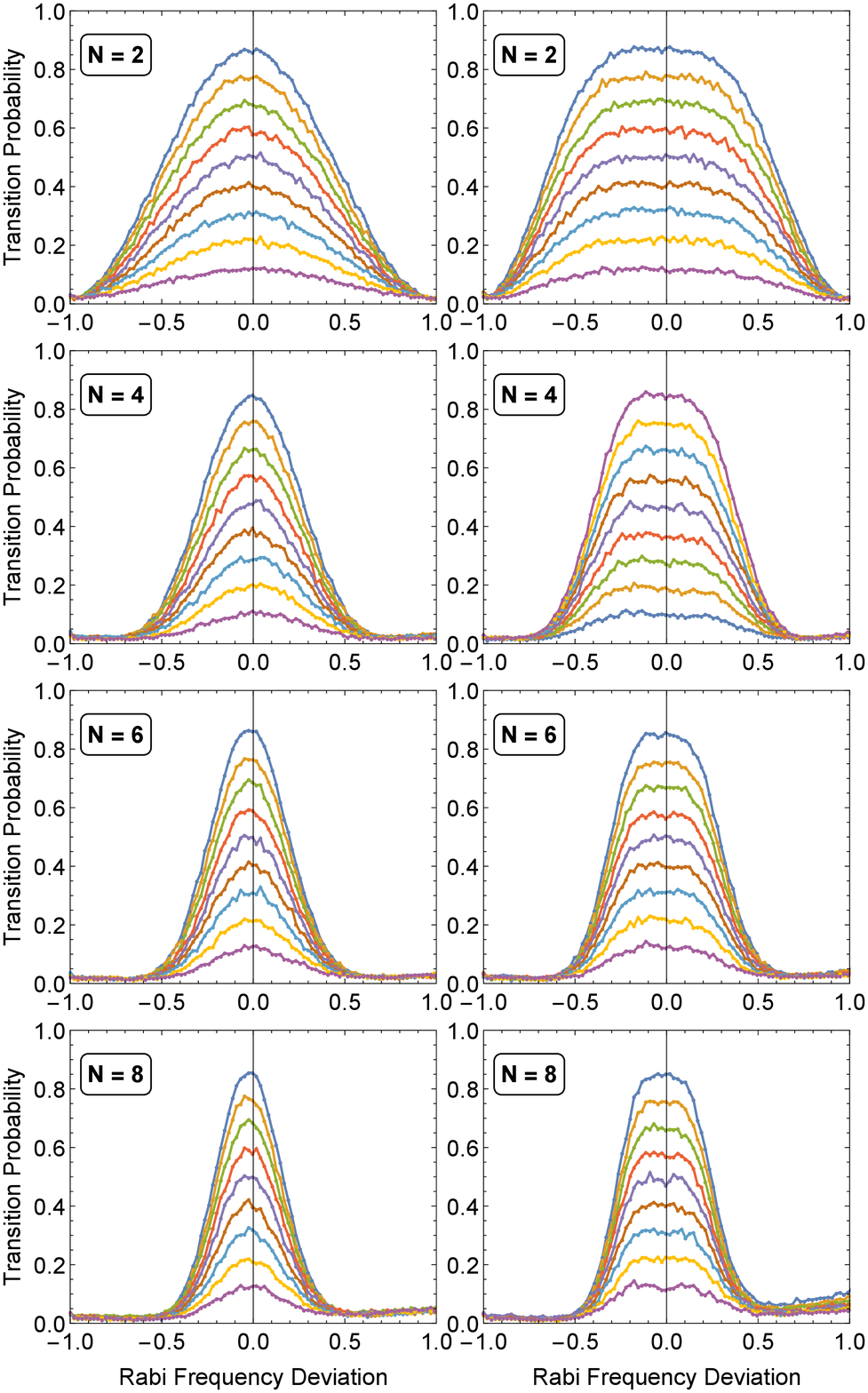}
	\caption{Transition probability induced by composite NB (left column) and PB (right column) theta pulses with lengths of $N$ and $2N$ respectively. The composite phases are given in Tables~\ref{Table:ThetaNB} and \ref{Table:ThetaPB}, and the transition probability is locked at levels $0.1,0.2,\ldots,0.9$.}
	\label{Fig:ThetaNBPB}
\end{figure}

\subsection{Broadband theta pulses}\label{sec-ThetaBB}

We start by testing the broadband theta pulses, derived in \cite{Torosov2019variable}. 
These sequences have the form
\be\label{ThetaSeq}
A_{\phi_1}B_{\phi_2}B_{\phi_3}\cdots B_{\phi_{N-1}}A_{\phi_N},
\ee
where $A$ is a nominal (i.e., for zero error) $\pi/2$ pulse and $B$ is a nominal $\pi$ pulse. 
The composite phases, for the chosen target probabilities, are given in Table.~\ref{Table:ThetaBB}. 
The excitation profiles are plotted in Fig.~\ref{Fig:ThetaBB}. 
They demonstrate first, the equidistant stepwise increase of the transition probability in the central part, and second, the broadening of the flat central part as the number of constituent pulses in the CP increases, signalling increasing robustness to Rabi frequency deviations.
The experimental results are, once again, in excellent agreement with the theoretical predictions, which can be found in \cite{Torosov2019variable}, omitted here for the sake of brevity.

\subsection{Narrowband and passband theta pulses}\label{sec-ThetaNB}

To derive narrowband $\theta$ CPs, the same type \eqref{ThetaSeq} of sequences have been used, as for the BB case. The composite phases, as derived in \cite{Torosov2020}, are given in Table~\ref{Table:ThetaNB}, and the corresponding experimentally measured profiles are plotted in Fig.~\ref{Fig:ThetaNBPB} (left column). 
As for the BB $\theta$ CPs in Fig.~\ref{Fig:ThetaBB}, the equidistant stepwise increase of the transition probability in the central part is clearly seen, indicating precise quantum control of the transition probability.
As predicted by theory, the central part squeezes as the number of constituent pulses in the CP increases, signalling increasing sensitivity to Rabi frequency deviations.
Again, the experimental results are in excellent agreement with the theoretical predictions, which can be found in \cite{Torosov2020} and are not shown here in order not to overcrowd the frames.


Finally, PB $\theta$ pulses can be obtained by ``twinning'' two narrowband $\pi/2$ CPs, where the second NB sequence is obtained by reversing the pulse order in the first one and its phases are shifted with the phase $\vartheta=2\arccos(\sqrt{p})$ \cite{Torosov2020}. 
The PB excitation profiles are plotted in Fig.~\ref{Fig:ThetaNBPB} (right column). 
As for the BB and NB CPs, the equidistant stepwise increase of the transition probability in the central part demonstrates the ability for precise quantum control of the transition probability.
Because these PB sequences are constructed from the NB ones, the excitation profiles squeeze as the number of constituent pulses in the CP increases.
Compared to the NB sequences (left column), the PB sequences (right column) generate a flat top of each profile, albeit not as broad as for the BB sequences in Fig.~\ref{Fig:ThetaBB}.
Hence they allow for both NB-like suppression of excitation outside the central region \textit{and} BB-like compensation for small errors in the central region.
This added benefit of the PB sequences compared to the BB and NB sequences comes at the expense of an increased length of the PB sequences by a factor of 2: note that the PB sequences in the right column of Fig.~\ref{Fig:ThetaNBPB} are twice longer (containing $2N$ pulses) than the corresponding NB sequences in the left column (containing $N$ pulses). 
Once again, we have verified that the experimental results are in excellent agreement with the theoretical predictions in \cite{Torosov2020}.


\section{Discussion and Conclusions}\label{sec-discussion}

In this work, we presented experimental results for a large number of composite pulse sequences, producing broadband, narrowband, and passband excitation profiles locked at any pre-selected target probability.
The experiments have been performed on a superconducting qubit, using the open-access cloud-based quantum-computing platform of IBM \cite{ibm_quantum}. 
All results demonstrate an excellent agreement between theory and experiment.
This agreement proves both the high quality of the transmon qubit, provided by IBM, as well as the control power of the composite pulse technique.

We sincerely hope that our results will stimulate \textit{first}, the use of composite pulses in superconducting quantum computing because they feature an unparalleled combination of great accuracy, precision, robustness and flexibility, and \textit{second}, to make use of the generous open-access resources of IBM Quantum to test other quantum control methods on real systems.

\acknowledgments
This work is supported by the European Commission's Horizon-2020 Flagship on Quantum Technologies project 820314 (MicroQC). We acknowledge the use of IBM Quantum services for this work. The views expressed are those of the authors, and do not reflect the official policy or position of IBM or the IBM Quantum team.

\appendix 

\section{Phases of BB, NB and PB $\theta$ composite pulses}

Here we present the phases of the composite $\theta$-pulses used in the experiments in this paper, which cannot be expressed by analytic formulas.
Table \ref{Table:ThetaBB} shows the phases for the BB CPs, Table \ref{Table:ThetaNB} the phases for the NB CPs, and Table \ref{Table:ThetaPB} the phases for the PB CPs.

\begin{widetext}

\begin{table*}[h]
	\begin{tabular}{|c|c|c|c|c|c|}
		\hline
		$p$ & 2 pulses & 3 pulses  & 4 pulses & 5 pulses & 6 pulses \\
		{} & $A_{0} A_{\phi_2}$ &
		$A_{0} B_{\phi_2} A_{\phi_3}$ &
		$A_{0} B_{\phi_2} B_{\phi_3} A_{\phi_4} $ &
		$A_{0} B_{\phi_2} B_{\phi_3} B_{\phi_4} A_{\phi_5}$ &
		$A_{0} B_{\phi_2} B_{\phi_3} B_{\phi_4} B_{\phi_5} A_{\phi_6}$ \\
		\hline
		{} & $\phi_2$  & $\phi_2,\phi_3$ &  $\phi_2,\phi_3,\phi_4$  &  $\phi_2,\phi_3,\phi_4,\phi_5$ & $\phi_2,\phi_3,\phi_4,\phi_5,\phi_6$ \\
		\hline
		$0.1$ & $0.7952 $ & $0.8204,1.4359$  & $2/3, 1.4618 , 0.7952$ & $0.5033,1.6110,1.1032,1.7861$ & $2/5, 8/5, 0.3952, 1.1952, 0.7952$\\
		$0.2$ & $0.7048 $ & $0.7952, 1.2952$  & $2/3, 1.3715, 0.7048$ & $0.4569, 1.5710, 1.185, 1.8467$ & $2/5, 8/5, 0.3048, 1.1048, 0.7048$\\
		$0.3$ & $0.6310 $ & $0.7778, 1.1866$  & $2/3, 1.2977, 0.6310$ & $0.4253, 1.5436, 1.2531, 1.9006$ & $2/5, 8/5, 0.2310, 1.0310, 0.6310$\\
		$0.4$ & $0.5641 $ & $0.7634, 1.0908$  & $2/3, 1.2308, 0.5641$ & $0.3991, 1.5209, 1.3153, 1.9510$ & $2/5, 8/5, 0.1641, 0.9641, 0.5641$ \\
		$0.5$ & $0.5 $    & $3/4, 1$  & $2/3, 7/6, 1/2$ & $3/8, 3/2, 11/8, 0$ & $2/5, 8/5, 1/10, 9/10, 1/2$\\
		$0.6$ & $0.4359 $ & $0.7366, 0.9092$  & $2/3, 1.1026, 0.4359$ & $0.3509, 1.4791, 1.4347, 0.0490$ & $2/5, 8/5, 0.0359, 0.8359, 0.4359$\\
		$0.7$ & $0.3690 $ & $0.7222, 0.8134$  & $2/3, 1.0357, 0.3690$ & $0.3247, 1.4564, 1.4969, 0.0994$ & $2/5, 8/5, 1.9689, 0.7689, 0.3689$\\
		$0.8$ & $0.2952 $ & $0.7048, 0.7048$  & $2/3, 0.9618 , 0.2952$ & $0.2931, 1.4291, 1.565, 0.1533$ & $2/5, 8/5, 1.8952, 0.6952, 0.2952$\\
		$0.9$ & $0.2048 $ & $0.6796, 0.5641$  & $2/3, 0.8715, 0.2048$ & $0.2467, 1.3890, 1.6468, 0.2139$ & $2/5, 8/5, 1.8048, 0.6048, 0.2048$\\
		\hline
	\end{tabular}
	\caption{
		Phases (in units $\pi$) of CPs which produce BB profiles with different target transition probability $p$. $A$ is a nominal (i.e., for zero error) $\pi/2$ pulse and $B$ is a nominal $\pi$ pulse.
	}
	\label{Table:ThetaBB}
\end{table*}

\begin{table*}[h]
	\begin{tabular}{|c|c|c|c|c|}
		\hline
		$p$ & 2 pulses & 4 pulses  & 6 pulses & 8 pulses \\
		{} & $A_{0} A_{\phi_2}$ &
		$A_{0} B_{\phi_2} B_{\phi_3} A_{\phi_4}$ &
		$A_{0} B_{\phi_2} B_{\phi_3} B_{\phi_4} B_{\phi_5} A_{\phi_6}$ &
		$A_{0} B_{\phi_2} B_{\phi_3} B_{\phi_4} B_{\phi_5}  B_{\phi_6} B_{\phi_7}A_{\phi_8}$\\
		\hline
		{} & $\phi_2$  & $\phi_2,\phi_3,\phi_4$ &  $\phi_2,\phi_3,\phi_4,\phi_5,\phi_6$  &  $\phi_2,\phi_3,\phi_4,\phi_5,\phi_6,\phi_7,\phi_8$  \\
		\hline
		$0.1$ & $0.7952 $ & $0.0769, 1.0257, 1.1026$  & $1.4150, 0.5716, 0.8499, 0.0064, 1.4214 $ & $1.2681, 0.5191, 0.4643, 1.5937, 1.5389, 0.7899, 0.0580$  \\
		$0.2$ & $0.7048 $ & $0.1108, 1.0373, 1.1481$  & $1.4316, 0.6075, 0.8012, 1.9772, 1.4087 $ & $1.2813, 0.5427, 0.4539, 1.6112, 1.5223, 0.7838, 0.0651$  \\
		$0.3$ & $0.6310 $ & $0.1386, 1.0469, 1.1855$  & $1.4379, 0.6284, 0.7646, 1.9551, 1.3930 $ & $1.2879, 0.5569, 0.4423, 1.6198, 1.5052, 0.7742, 0.0621$  \\
		$0.4$ & $0.5641 $ & $0.1639, 1.0557, 1.2196$  & $1.4400, 0.6430, 0.7330, 1.9360, 1.3760 $ & $1.2917, 0.5672, 0.4302, 1.6248, 1.4879, 0.7633, 0.0551$ \\
		$0.5$ & $0.5 $    & $0.1881, 1.0644, 1.2525$  & $1.4396, 0.6541, 0.7038, 1.9182, 1.3579 $ & $1.2939, 0.5752, 0.4177, 1.6277, 1.4702, 0.7515, 0.0454$ \\
		$0.6$ & $0.4359 $ & $0.2124, 1.0732, 1.2857$  & $1.4374, 0.6629, 0.6752, 1.9008, 1.3382 $ & $1.2948, 0.5818, 0.4043, 1.6291, 1.4516, 0.7386, 0.0334$ \\
		$0.7$ & $0.3690 $ & $0.2379, 1.0827, 1.3207$  & $1.4334, 0.6702, 0.6460, 1.8828, 1.3162 $ & $1.2947, 0.5874, 0.3896, 1.6291, 1.4314, 0.7241, 0.0187$ \\
		$0.8$ & $0.2952 $ & $0.2661, 1.0936, 1.3597$  & $1.4274, 0.6763, 0.6142, 1.8630, 1.2904 $ & $1.2934, 0.5922, 0.3727, 1.6277, 1.4081, 0.7069, 0.0003$ \\
		$0.9$ & $0.2048 $ & $0.3009, 1.1075, 1.4083$  & $1.4183, 0.6813, 0.5755, 1.8385, 1.2568 $ & $1.2906, 0.5965, 0.3508, 1.6240, 1.3784, 0.6843, 1.9749$ \\
		\hline
	\end{tabular}
	\caption{
		Phases (in units $\pi$) of CPs which produce NB profiles with different target transition probability $p$. $A$ is a nominal (i.e., for zero error) $\pi/2$ pulse and $B$ is a nominal $\pi$ pulse.
	}
	\label{Table:ThetaNB}
\end{table*}


\begin{table*}[h]
	\begin{tabular}{| c | c | c | c | c | c |}
		\hline
		$p$ & $\vartheta$ & 4 pulses & 8 pulses  & 12 pulses & 16 pulses \\
		 &  & $A_{0} A_{\phi_2} A_{\phi_2+\vartheta} A_{\vartheta}$ &
		\begin{tabular}{c}
		     $A_{0} B_{\phi_2} B_{\phi_3} A_{\phi_4}$  \\
		     $A_{\phi_4+\vartheta} B_{\phi_3+\vartheta} B_{\phi_2+\vartheta} A_{\vartheta}$
		\end{tabular}
        & 
		\begin{tabular}{c}
		$A_{0} B_{\phi_2} B_{\phi_3} B_{\phi_4} B_{\phi_5} A_{\phi_6}$  \\
	  $A_{\phi_6+\vartheta} B_{\phi_5+\vartheta} B_{\phi_4+\vartheta}$ \\ $B_{\phi_3+\vartheta} B_{\phi_2+\vartheta} A_{\vartheta}$		      
		\end{tabular}
        &
        \begin{tabular}{c}
        $A_{0} B_{\phi_2} B_{\phi_3} B_{\phi_4} B_{\phi_5}  B_{\phi_6} B_{\phi_7}A_{\phi_8}$      \\
		  $A_{\phi8+\vartheta} B_{\phi_7+\vartheta} B_{\phi_6+\vartheta} B_{\phi_5+\vartheta}$ \\
		  $B_{\phi_4+\vartheta}  B_{\phi_3+\vartheta} B_{\phi_2+\vartheta}A_{\vartheta}$
        \end{tabular}
 \\
		\hline
		{} & {} & $\phi_2$  & $\phi_2,\phi_3,\phi_4$ &  $\phi_2,\phi_3,\phi_4,\phi_5,\phi_6$  &  $\phi_2,\phi_3,\phi_4,\phi_5,\phi_6,\phi_7,\phi_8$  \\
		\hline
		$0.1$ & 0.7952 &  &  &  &  \\
		$0.2$ & 0.7048 &  &  &  &    \\
		$0.3$ & 0.6310 &  &  &  &  \\
		$0.4$ & 0.5641 & $0.5$ & $0.1881, 1.0644, 1.2525$  & $1.4396, 0.6541, 0.7038, $ & $1.2939, 0.5752, 0.4177, 1.6277, $ \\
		$0.5$ & 0.5 &  &  & $ 1.9182, 1.3579 $ & $1.4702, 0.7515, 0.0454$  \\
		$0.6$ & 0.4359 &  &  &  &  \\
		$0.7$ & 0.3690 & in all cases & in all cases & in all cases & in all cases \\
		$0.8$ & 0.2952 &  &  &  &  \\
		$0.9$ & 0.2048 & &  &  &  \\
		\hline
	\end{tabular}
	\caption{
		Phases (in units $\pi$) of CPs which produce PB profiles with different target transition probability $p$. 
		The precise value of $\vartheta$ is $\vartheta=2\arccos(\sqrt{p})$. $A$ is a nominal (i.e., for zero error) $\pi/2$ pulse and $B$ is a nominal $\pi$ pulse.
	}
	\label{Table:ThetaPB}
\end{table*}

\end{widetext}



\end{document}